\documentclass[twocolumn,aps,pra,superscriptaddress]{revtex4}
\usepackage{amsmath,amssymb,mathrsfs,bm}
\usepackage{graphicx,dcolumn,times}
\usepackage{xcolor}
\usepackage{lipsum}
\usepackage{makecell}
\begin{document}

\title{ Microwave electrometry via electromagnetically induced absorption in cold Rydberg atoms}

\author{Kai-Yu Liao} \email{kaiyu.liao@m.scnu.edu.cn}
\affiliation{Guangdong Provincial Key Laboratory of Quantum Engineering and Quantum Materials, GPETR Center for Quantum Precision Measurement and SPTE, South China Normal University, Guangzhou 510006, China}

\author{Hai-Tao Tu} 
\affiliation{Guangdong Provincial Key Laboratory of Quantum Engineering and Quantum Materials, GPETR Center for Quantum Precision Measurement and SPTE, South China Normal University, Guangzhou 510006, China}

\author{Shu-Zhe Yang}
\affiliation{Guangdong Provincial Key Laboratory of Quantum Engineering and Quantum Materials, GPETR Center for Quantum Precision Measurement and SPTE, South China Normal University, Guangzhou 510006, China}

\author{Chang-Jun Chen}
\affiliation{Guangdong Provincial Key Laboratory of Quantum Engineering and Quantum Materials, GPETR Center for Quantum Precision Measurement and SPTE, South China Normal University, Guangzhou 510006, China}

\author{Xiao-Hong Liu}
\affiliation{Guangdong Provincial Key Laboratory of Quantum Engineering and Quantum Materials, GPETR Center for Quantum Precision Measurement and SPTE, South China Normal University, Guangzhou 510006, China}

\author{Jie Liang}
\affiliation{Guangdong Provincial Key Laboratory of Quantum Engineering and Quantum Materials, GPETR Center for Quantum Precision Measurement and SPTE, South China Normal University, Guangzhou 510006, China}

\author{Xin-Ding Zhang}
\affiliation{Guangdong Provincial Key Laboratory of Quantum Engineering and Quantum Materials, GPETR Center for Quantum Precision Measurement and SPTE, South China Normal University, Guangzhou 510006, China}

\author{Hui Yan} \email{yanhui@scnu.edu.cn}
\affiliation{Guangdong Provincial Key Laboratory of Quantum Engineering and Quantum Materials, GPETR Center for Quantum Precision Measurement and SPTE, South China Normal University, Guangzhou 510006, China}

\affiliation{ Frontier Research Institute for Physics,
South China Normal University, Guangzhou 510006, China}

\author{Shi-Liang Zhu} \email{slzhu@nju.edu.cn}
\affiliation{National Laboratory of Solid State Microstructures, School of Physics, Nanjing University, Nanjing 210093, China}
\affiliation{Guangdong Provincial Key Laboratory of Quantum Engineering and Quantum Materials, GPETR Center for Quantum Precision Measurement and SPTE, South China Normal University, Guangzhou 510006, China}
\affiliation{ Frontier Research Institute for Physics,
South China Normal University, Guangzhou 510006, China}


\begin{abstract}
The atom-based traceable standard for microwave electrometry shows promising advantages by enabling stable and uniform measurement. Here we theoretically propose and then experimentally realize an alternative direct International System of Units (SI)-traceable and self-calibrated method for measuring a microwave electric field strength based on electromagnetically induced absorption (EIA) in cold Rydberg atoms. Comparing with the method of electromagnetically induced transparency, we show that the equivalence relation between microwave Rabi frequency and Autler-Townes splitting is more valid and is even more robust against the experimental parameters in the EIA's linear region. Furthermore, a narrower linewidth of cold Rydberg EIA enables us to realize a direct SI-traceable microwave-electric-field measurement as small as $\sim$100 $\mu\mathrm{\!V} \mathrm{cm}^{\!-\!1}$.

\end{abstract}
\maketitle


\section{Introduction}

A stated goal of metrology organizations is to make all measurements quantum traceable in the International System of Units (SI), and the science of measurement has been  changing rapidly due to the SI redefinition since 2018 \cite{Stock2018, NIST2018}. On account of the shift towards fundamental physical constants, the role of basic standards must change, which includes the microwave (MW) electric field  and power.
The reproducibility, accuracy and stability of atom-based metrologies significantly outperform conventional methods due to the stability and uniformity of the atomic properties \cite{force, magneto, magneto2, clock, clock2, gravity, atom, sensing, nvcenter}.  Pioneering experiments \cite{shaffer1,cat} have demonstrated that Rydberg atoms can be used to measure the MW \cite{shaffer1} or static \cite{cat}  electric field with higher accuracy, sensitivity and stability then the traditional antenna-based method \cite{micrometer, review}. This novel approach has since been exploited extensively for antenna calibration \cite{shaffer2, strong, calibration, jianming}, signal detection \cite{signal, zhenfei,linjie, adams}, subwavelength imaging \cite{imaging}, and terahertz sensing \cite{THz}.


The Rydberg-atom-based MW electrometry utilizes the phenomena of electromagnetically induced transparency (EIT) and Autler-Townes splitting (ATS) \cite{eit, eit&ats, eit&ats2}.  In the method, the Rabi frequency $\Omega_{\!M\!W}$ of the MW atomic transition is considered to be equal to the ATS ($2\pi\Delta\!f$) in the EIT spectrum:
\begin{equation}
\label{Rabi_AT}
\Omega_{\!M\!W} = 2\pi\Delta\!f.
\end{equation}
Then the MW electric field $|E|=2\pi\hbar  \Delta\!f/\mu$ \cite{shaffer1}. This type of measurement of the MW electric field is just a goal of metrology organizations: It is a direct
SI-traceable, self-calibrated measurement in that it is related to Planck's constant (which is an SI-defined quantity) and only requires a frequency measurement ($\Delta\!f$, which can be measured very accurately and is calibrated to the hyperfine atomic structure). The atomic dipole moment $\mu$ is a parameter which can be calculated accurately \cite{shaffer1,uncertainty}. This method can be further developed for a direct SI-traceable measurement for power metrology \cite{Holloway2018}.


Various aspects of the uncertainties of this measurement approach have been investigated \cite{shaffer1, cell}. In particular, the validity of Eq. (1) is an essential aspect of uncertainties of this approach \cite{uncertainty}. However, for hot atoms, though the ATS is regularly measured via a double-peak fit, it is a challenge to examine the validity
as a result of the inhomogeneous broadening in the vapor-cell spectrum \cite{rydeit, calibration}. Furthermore, the accuracy and resolution of direct SI-traceable MW measurements are closely related to the EIT linewidth. The full width at half-maximum of Rydberg EIT on record for room-temperature atomic vapor is about several megahertz owing to the Doppler mismatching in the three-level cascade system with the dual-wavelength lasers \cite{uncertainty}. So the low bound of the direct SI-traceable MW-electric-field strength is limited to around 5 $\mathrm{m\!V}\mathrm{cm}^{-1}\!$.

Motivated by the narrower linewidth and lower dephasing rates of cold atoms \cite{wenhui2, zhao}, here  we theoretically propose and then experimentally  realize an alternative direct SI traceable MW electrometry with cold Rydberg atoms using electromagnetically induced  absorption (EIA). Cold atoms are employed here to obtain the atomic resonances with a subnatural linewidth, and their solvable density matrix allows us to analyze the relationship between the MW Rabi frequency and ATS clearly \cite{zhao, dds, wenhui}.   We find that, besides the MW Rabi frequency, the ATS depends on various system parameters (coupling Rabi frequency, atomic dephasing rates, etc.) in the ordinary EIT regime, and thus  the equivalence relation given in Eq. (\ref{Rabi_AT}) breaks down. However, in the EIA regime, where the excited state is adiabatically eliminated with a large single-photon detuning, the relation (1) remains valid and is robust against variations of the experimental parameters. We then demonstrate an alternative approach for direct SI-traceable MW-electric-field measurement.  Using the narrower EIA signal, we achieve a direct SI-traceable MW-electric-field measurement as small as 100 $\mu\mathrm{\!V}\mathrm{cm}^{-\!1}$ in free space, about $\frac{1}{50}$ the low bound achievable by vapor-cell EIT method.  Our work reports a highly accurate and sensitive direct SI-traceable measurement, and therefore constitutes a major step towards the atom-based standard of MW electrometry.


The paper is organized as follows. In Sec.II, we introduce the experimental setup. Section III introduces the theoretical model. In Sec. IV, we discuss the experimental results. A brief discussion is given in Sec. V. We summarize in Sec. VI. In the Appendix, we present the principal sources of the uncertainties in our scheme.

\begin{figure}[t]
\begin{center}
\label{Setup}
\includegraphics[width=8cm]{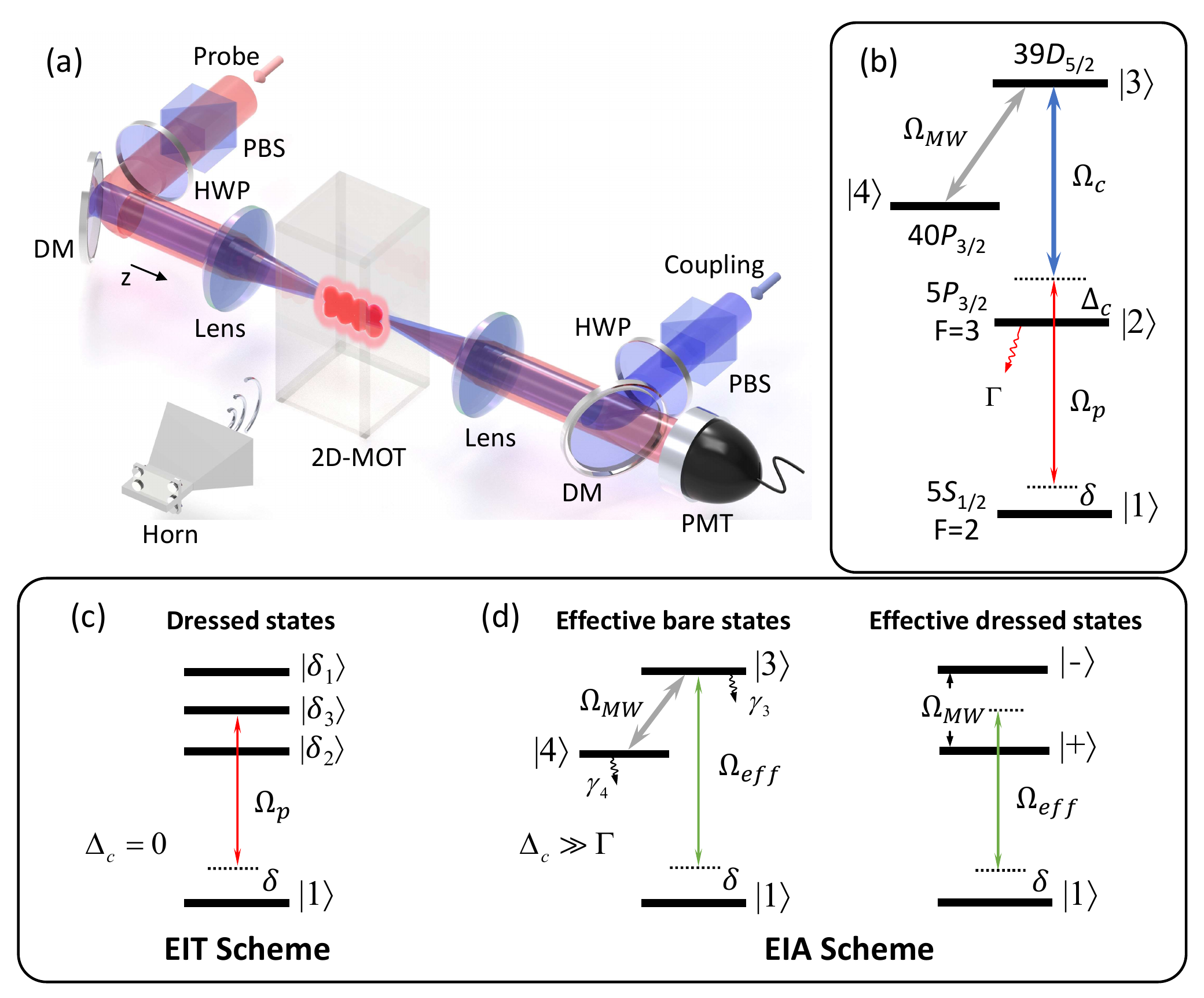}
\caption{ (a) Schematic of the experimental setup. The probe and coupling lasers overlap inside a two-dimensional magneto-optical trap (2D-MOT) and a horn antenna is placed in the far-field limit. (b) Relevant energy levels of $^{87}$Rb atoms. A weak 780 nm laser, a strong 480-nm laser, and a MW electric field ($\sim$36.9 GHz) are coupled to three electric dipole transitions $|1\rangle\!\rightarrow\!|2\rangle$, $|2\rangle \!\rightarrow\!|3\rangle$, and $|3\rangle\!\rightarrow\!|4\rangle$, respectively.
(c) Four-level EIT scheme
in the decaying-dressed state basis. (d) Effective three-level system for the EIA scheme in the basis of both the bare states and the decaying dressed states. With the adiabatic elimination of the excited state $|2\rangle$, the dressed-state frequency separation is equal to the MW Rabi frequency $\Omega_{\!M\!W}$. }
\end{center}
\end{figure}

\section{Experimental setup} The experimental setup and the MW-driven four-energy-level system are shown in Fig.1.  The experiment cycle has a period of 5 ms, in which 0.45 ms is for the atom preparation and 0.5 ms is for measurement window. 
A cigar-shaped $^{87}$Rb cloud, with longitudinal length $L$ = 2.2 cm, temperature $T\!\sim$100 $\mu$K and atomic density $N\!\sim$10$^{\mathrm{10}}\mathrm{cm}^\mathrm{-3}$, is prepared in the hyperfine state $|1\rangle = | 5S_{1 / 2}, F=2 \rangle$ within 4.5 ms. Along the $z$ direction, the cloud has an optical depth ($O\!D$) up to 140 in the $|1\rangle \rightarrow\! |2\rangle = |5P_{3 / 2}, F=3\rangle$ transition \cite{mot}. The counter propagating vertically polarized probe and coupling laser beams are focused to $1 /{e}^{2}$ radii of 50 and 85 $\mu$m, respectively. The frequencies of the probe and coupling lasers are locked to a high-finesse temperature stabilized Fabry-Perot cavity through the Pound-Drever-Hall technique \cite{FP}. The stray magnetic field is compensated by three pairs of Helmholtz coils down to 10 mG \cite{memory}, and the glass cell is shielded with MW absorbers. The vertically polarized MW is emitted from a horn antenna, driven by a MW generator (R\&S SMF100A), and propagates perpendicularly to the probe and coupling beams. The MW with a frequency of 36.8961 GHz is resonant with two adjacent Rydberg energy levels \cite{mack}, $|3\rangle = 39 D_{5 / 2} \leftrightarrow |4\rangle = 40 P_{3 / 2}$, with a radial transition dipole moment $\mu$ = 1926 ${ea}_{0}$. The coupling laser is on during the whole measurement window. The power of the probe laser ($P_{0} =$ 0.8 nW), is kept constant during the experiment, which corresponds to a Rabi frequency of $\Omega_{p}/2\pi$ = 0.4 MHz. The frequency of the probe laser is scanned around the $|1\rangle\!\rightarrow\!|3\rangle$ two-photon resonance from -20 to 20 MHz within 100 $\mu$s using an acousto-optic modulator at the beginning of the 0.5 ms measurement window. We scan only one time during this measurement window in the present experiments, although scanning five times is in principle possible.
The transmission spectrum $P_{\!t}$ of the probe laser, described by $P_{t}=P_{0}\exp(-\alpha L)$, with the absorption coefficient $\alpha\!\propto\!N\!\operatorname{Im}(\varrho_{p})$ and $\varrho_{p}$ the atomic polarization
given in Eq. (\ref{rho21}) or (\ref{rho31}), is recorded by a photomultiplier tube (PMT) (Hamamatsu, H10720-20).


\section{Theoretical model}

Our system is schematically shown in Fig. 1(b); It is a four-level atom interacting with the MW field and coupling and probe lasers.
Using the resonant pole method, Rawat \textit{et al.} investigated the interference in a MW-driven four-level cascade atomic vapor and made a distinction between double EIT and double ATS \cite{dds}. Here we apply this theoretical method to study the case in cold Rydberg gases and introduce the EIA ATS domain under the condition of large coupling detuning. We will introduce a detailed method to calculate the atomic polarizations in this section.

\subsection{Decaying-dressed states of four-level atom}
We consider the rubidium cascade four-level system illustrated in Fig. 1(b). Because of the low atomic density and very small Rydberg populations, the interactions between the Rydberg atoms are weak and thus the model is described by effective dephasing rates in the single-atom picture. In the electric dipole and rotating-wave approximation, the Hamiltonian for a four-level atom interacting with the three external fields is given by
\begin{equation}
\begin{aligned}
\label{Hamiltonian}
H =&-\hbar\left[(\delta-\Delta_{c}) A_{22}+ \delta A_{33}+ (\delta-\Delta_{\!M\!W}) A_{44}\right]\\
&+\frac{\hbar}{2}(\Omega_{p} A_{21}+\Omega_{c} A_{32}+\Omega_{\!M\!W}A_{43}+\mathrm{H.c.}),
\end{aligned}
\end{equation}
where $A_{i j}=|i\rangle\langle j|$ are atomic transition operators and $\Omega_{i}=\mu_{i} E_{i} / \hbar$ (with $i=p,c,{MW}$) are the Rabi frequency of the incident fields.

To account for the decay and dephasing, we model the time evolution of the atomic system by a Lindblad master equation for the density operator $\varrho$
\begin{equation}
\partial_{t} \varrho=-\frac{i}{\hbar}\left[H, \varrho\right]+\mathcal{L}_{\gamma}\varrho+\mathcal{L}_{\text{deph}}\varrho.
\end{equation}
In this equation, the term $\mathcal{L}_{\gamma}\varrho$ representing the spontaneous emission of the upper levels  is described by standard Lindblad decay terms
\begin{equation}
\begin{aligned}
\mathcal{L}_{\gamma} \varrho=&-\frac{\Gamma}{2}\left(A_{22} \varrho+\varrho A_{22}-2 A_{12} \varrho A_{12}^{\dagger}\right) \\
&-\frac{\gamma}{2}\left(A_{33} \varrho+\varrho A_{33}-2 A_{23} \varrho A_{23}^{\dagger}\right)\\
&-\frac{\gamma^{\prime}}{2}\left(A_{44}\varrho+\varrho A_{44}-2 A_{14} \varrho A_{14}^{\dagger}\right).
\end{aligned}
\end{equation}
Here $\Gamma$ is the decay rate from the excited state $|2\rangle$ to the ground state $|1\rangle$. The Lindblad terms with $\gamma$ and $\gamma^{\prime} \ll \Gamma$ account for the decay of the meta-stable Rydberg states $|3\rangle$ and $|4\rangle$, respectively. The last term in Eq. (\ref{Hamiltonian}) represents the dephasing of atomic coherence owing to atomic collisions, a finite laser linewidth, and dipole-dipole interactions between Rydberg atoms,
\begin{equation}
\begin{aligned}
\mathcal{L}_{\text {deph }} \varrho=&-\gamma_{d}\left(A_{33} \varrho+\varrho A_{33}-2 A_{33} \varrho A_{33}\right) \\ &-\gamma_{d}^{\prime} \left(A_{44} \varrho+\varrho A_{44}-2 A_{44} \varrho A_{44}\right).
\end{aligned}
\end{equation}

We obtain the steady-state solutions of the master equation by applying the weak probe condition. The ground-state approximation is equivalent to letting $\varrho_{11}\simeq 1$, $\varrho_{22}=\varrho_{33}=\varrho_{44}=0$, and $\varrho_{23}=\varrho_{24}=\varrho_{34}=0$. The off-diagonal element corresponding to the probe transition is obtained as
\begin{equation}
\label{rho21V1}
\varrho_{21}=\frac{\Omega_{p}/2}{\delta-\Delta_{c}-i \gamma_{2}+\frac{\Omega_{c}^{2}(\delta-\Delta_{\!M\!W}-i \gamma_{4})}{\Omega_{\!M\!W}^{2}-4(\delta-i \gamma_{3})(\delta-\Delta_{\!M\!W}-i \gamma_{4})}},
\end{equation}
where $\gamma_{2} = \Gamma/2$, $\gamma_{3} = \gamma/2+\gamma_{d}$, and $\gamma_{4} = \gamma^{\prime}/2+\gamma_{d}^{\prime}$.
We can further rewrite Eq. (\ref{rho21V1}) as
\begin{equation}
\label{rho21}
\varrho_{21}(\delta)=\frac{\Omega_{p}}{2}\frac{d_{3}d_{4}-\Omega_{\!M\!W}^{2}/4} {d_{2}d_{3}d_{4}-d_{2}\Omega_{\!M\!W}^{2}/4-d_{4}\Omega_{c}^{2}/4},
\end{equation}
where $\Omega_{i}$ are the Rabi frequencies of incident fields, and the complex detunings $d_2=\delta-\Delta_{c}-i \gamma_{2}$, $d_3=\delta-i \gamma_{3}$, and $d_4=\delta-\Delta_{\!M\!W}-i \gamma_{4}$ with $\gamma_{i}$ being the total dephasing rates of $|i\rangle$ \cite{dds}.
In experiments, the total dephasing rates $\gamma_{3}$ and $\gamma_{4}$ are free parameters and their values are obtained from a fit to experimental data, while a common parameter is $\Gamma =2 \pi \times$ 6 MHz.

The linear response is a complicated expression to analyze, and thus we rewrite it as a function of the complex variable $\delta$:
\begin{equation}
\label{rho21V2}
\varrho_{21}(\delta)=\frac{\Omega_{p}}{2}\frac{\left(\delta-i \gamma_{3}\right)\left(\delta-\Delta_{\!M\!W}-i \gamma_{4}\right)-\Omega_{\!M\!W}^{2}/4}
{\left(\delta-\delta_{1}\right)\left(\delta-\delta_{2}\right)\left(\delta-\delta_{3}\right)}.
\end{equation}
Equation (\ref{rho21V2}) indicates that the linear response has three resonant poles:
\begin{equation}
\delta_{1}=\frac{1}{3}\left(D+\sqrt[3]{2} \frac{L_{1}}{L_{3}}-\frac{L_{3}}{\sqrt[3]{2}}\right),
\end{equation}
\begin{equation}
\delta_{2}=\frac{1}{3}\left(D-\frac{(1+i \sqrt{3}) L_{1}}{\sqrt[3]{4} L_{3}}+\frac{(1-i \sqrt{3}) L_{3}}{2 \sqrt[3]{2}}\right),
\end{equation}
\begin{equation}
\delta_{3}=\frac{1}{3}\left(D-\frac{(1-i \sqrt{3}) L_{1}}{\sqrt[3]{4} L_{3}}+\frac{(1+i \sqrt{3}) L_{3}}{2 \sqrt[3]{2}}\right).
\end{equation}
For the sake of simplicity, the above variables $D, L_1$, $L_2$, and $L_3$ are defined as follows:
\begin{equation}
D=-(d_{2}+d_{3}+d_{4}),
\end{equation}
\begin{equation}
L_{1}=-d_{2}^{2}+d_{2} d_{3}-d_{3}^{2}+d_{2} d_{4}+d_{3} d_{4}-d_{4}^{2}-\frac{3}{4} \Omega_{c}^{2}-\frac{3}{4} \Omega_{\!M\!W}^{2},
\end{equation}
\begin{equation}
\begin{aligned}
L_{2} = & 2 d_{2}^{3}-3 d_{2}^{2} d_{3}-3 d_{3}^{2} d_{2}-3 d_{2}^{2} d_{4}-3 d_{3}^{2} d_{4}-3 d_{4}^{2} d_{2} \\&-3 d_{4}^{2} d_{3}+2 d_{3}^{3}+12 d_{2} d_{3} d_{4}+2 d_{4}^{3}+\frac{9}{4}d_{2} \Omega_{c}^{2}+\frac{9}{4} d_{3} \Omega_{c}^{2}\\&+\frac{9}{4} d_{3} \Omega_{\!M\!W}^{2} +\frac{9}{4} d_{4} \Omega_{\!M\!W}^{2}-\frac{9}{2} d_{4} \Omega_{c}^{2}-\frac{9}{2} d_{2} \Omega_{\!M\!W}^{2},
\end{aligned}
\end{equation}
\begin{equation}
L_{3}=(L_{2}+\sqrt{4 L_{1}^{3}+L_{2}^{2}})^{1 / 3}.
\end{equation}
Here we have introduced the complex detunings  $d_{2}=\delta-\Delta_{c}-i \gamma_{2}$, $d_3=\delta-i \gamma_{3}$ and $d_4=\delta-\Delta_{\!M\!W}-i \gamma_{4}$.

 The slowly varying amplitude $\varrho_{21}$ can be further expressed as a superposition of three resonant responses associated with the transitions from the ground state to the corresponding decaying dressed states
\begin{equation}
\varrho_{21}(\delta) =\frac{\Omega_{p}}{2}\sum_{i=1}^{3}\frac{ S_{i}}{\delta-\delta_{i}},
\end{equation}
where
\begin{equation}
S_{1} =-\frac{d_{3} d_{4}+d_{3} \delta_{1}+d_{4} \delta_{1}+\delta_{1}^{2}-\Omega_{\!M\!W}^{2}/4}{\left(\delta_{1}-\delta_{2}\right)\left(\delta_{1}-\delta_{3}\right)},
\end{equation}
\begin{equation}
S_{2} =\frac{d_{3} d_{4}+d_{3} \delta_{2}+d_{4} \delta_{2}+\delta_{2}^{2}-\Omega_{\!M\!W}^{2}/4}{\left(\delta_{1}-\delta_{2}\right)\left(\delta_{2}-\delta_{3}\right)},
\end{equation}
and
\begin{equation}
S_{3} =\frac{d_{3} d_{4}+d_{3} \delta_{3}+d_{4} \delta_{3}+\delta_{3}^{2}-\Omega_{\!M\!W}^{2}/4}{\left(\delta_{1}-\delta_{3}\right)\left(-\delta_{2}+\delta_{3}\right)}.
\end{equation}
It has three poles representing the resonant responses to the probe field.  The imaginary part of $\varrho_{21}$ gives rise to absorption of the probe field by the cold ensemble, and thus the probe spectrum can be decomposed into three resonant terms
\begin{equation*} P_{\!t}/P_{\!0}=\prod_{i=1}^{3} R_{i}(\delta),\end{equation*} where
\begin{equation*}
R_{i}(\delta) =\exp \{-(O\!D \Gamma/2) \operatorname{Im}[{S_{i}}/({\delta-\delta_{i}})]\}.
\end{equation*}
The three poles attribute to the decaying dressed states of the four-level system with the level shifts and dephasing rates defined by the real and imaginary parts of $\delta_{i}$, respectively.

To have a physical understanding of the probing transmission, we discuss our system with the decaying dressed-state approach for two cases: EIT with $\Delta_c=0$ and EIA with large detuning $\Delta_c$.  Under the condition $\Delta_c=0$, the decaying dressed states arise from the interaction between the usual dressed states with the eigenenergies 0 and $\pm\!\sqrt{\Omega_{c}^{2}+\Omega_{\!M\!W}^{2}}/2$ and three reservoirs with dephasing rates $\gamma_{2}$, $\gamma_{3}$ and $\gamma_{4}$. The resonant responses are shown in Fig. 1(c), which is associated with the transitions from the ground state to corresponding decaying dressed states with level shifts and dephasing rates given by Re($\delta_i$) and Im($\delta_i$), respectively. These level shifts demonstrate that the splitting of the EIT peaks depends on a number of factors, including the Rabi frequencies of coupling and MW fields as well as the dephasing rates. Thus it implies that the relation (\ref{Rabi_AT}) should be carefully checked.

This analytical method gives clear insight into the nature of the interference of the MW-driven four-level system: One can observe EIT as a result of destructive interference if the resonance $R_{i} > 1$, whereas ATS is observed as a result of constructive interference.

\subsection{EIA-ATS regime}

For $\Delta_{c} \gg \Omega_{c}, \Gamma$, the excited state $|2\rangle$ can be adiabatically eliminated, as shown in Fig. 1(d). Under this condition, the Hamiltonian given in Eq. (\ref{Hamiltonian}) can be reduced to an effective three-level system represented in the subspace $\{|1\rangle,|3\rangle,|4\rangle\}$,
\begin{equation}
H_{e\!f\!f}=\frac{\hbar}{2}\left(\begin{array}{ccc}{0} & {\Omega_{e\!f\!f}} & {0} \\ {\Omega_{e\!f\!f}} & {-2 \left(\delta+\Delta_{\!A\!C}\right)} & {\Omega_{\!M\!W}} \\ {0} & {\Omega_{\!M\!W}} & {-2\left(\delta-\Delta_{\!M\!W}\right)}\end{array}\right),
\end{equation}
where the effective Rabi frequency $\Omega_{e\!f\!\!f}= \Omega_{p} \Omega_{c} /\left(2 \Delta_{c}\right)$ and AC Stark shift $\Delta_{\!A\!C}= (\Omega_{p}^{2}+\Omega_{c}^{2}) /\left(4 \Delta_{c}\right)$. To account for the electromagnetically induced two-photon absorption, we obtain the off-diagonal element $\varrho_{31}$ as
\begin{equation}
\varrho_{31}(\delta)=\frac{\Omega_{e\!f\!f}}{2}\frac{d_{4}}{d_{4}\left(d_{3}+\Delta_{\!A\!C}\right)-\Omega_{\!M\!W}^{2} /4}.
\end{equation}

The effective atomic coherence $\varrho_{31}$ has two absorption poles:
\begin{equation}
\delta_{\pm}=\frac{1}{2}[-\Delta_{\!A\!C}+i\left(\!\gamma_{3}\!+\!\gamma_{4}\right)\!\pm\!\sqrt{\!\Omega_{\!M\!W}^{2}\!+
\!\left(\Delta_{\!A\!C}\!-\!i\gamma_{3}\!+\!i\gamma_{4}\right)^{2}}].
\end{equation}
We rewrite the atomic coherence as a superposition of two resonant responses associated with the transitions from the ground state to the two decaying-dressed states:
\begin{equation}
\varrho_{31}=\frac{\Omega_{e\!f\!f}}{2}\left(\frac{ S_{+}}{\delta-\delta_{+}}+\frac{ S_{-}}{\delta-\delta_{-}}\right)
\end{equation}
with the strengths
\begin{equation}
S_{\pm}=\pm\frac{\delta_{\pm}-i \gamma_{{4}}}{\delta_{+}-\delta_{-}}.
\end{equation}
For $\Omega_{\!M\!W} \gg |\gamma_{3}-\gamma_{4}|, \Delta_{\!A\!C}$, we have $\delta_{\pm} \simeq [\pm\Omega_{\!M\!W}+i(\gamma_{3}+\gamma_{4})]/2$ and $S_{\pm}=1/2$.
 In this case, a strong-coupling field induces the probe absorption near the two-photon resonance via the $|1\rangle\!\rightarrow\!|3\rangle$ transition \cite{EIA}.
The EIA polarization has two absorption poles, which makes it a superposition of two resonances only associated with the MW Rabi frequency and Rydberg dephasing. In the effective three-level system, the threshold Rabi frequency of the MW field for the transition between EIT and ATS is $\left|\gamma_{3}-\gamma_{4}\right|$ \cite{eit&ats, eit&ats2}. As the Rydberg dephasing rates $\gamma_{3}$ and $\gamma_{4}$ are very close, $\Omega_{\!M\!W}\!\gg\!\left|\gamma_{3}\!-\!\gamma_{4}\right|, \Delta_{\!A\!C}$ (AC Stark shift), the EIA polarization can be written as the sum of two equal-width Lorentzians shifted from the two-photon resonance by $\pm \Omega _{M\!W}\!/2$,
\begin{equation}
\begin{aligned}
\label{rho31}
{\varrho_{31}(\delta)}\simeq &\frac{{{\Omega _{e\!f\!\!f}}}}{4}[\frac{1}{{\left( {\delta+{\Omega _{\!M\!W\!}}/2} \right) - i\left( {{\gamma _{3}} + {\gamma_{4}}} \right)/2}} \\ &+ \frac{1}{{\left( {\delta-\Omega_{\!M\!W\!}/2} \right) - i\left( {{\gamma_{3}} + {\gamma_{4}}} \right)/2}}],
\end{aligned}
\end{equation}
and thereby it is referred to as the EIA ATS regime.

\begin{figure*}[t]
\begin{center}
\includegraphics[width=16cm]{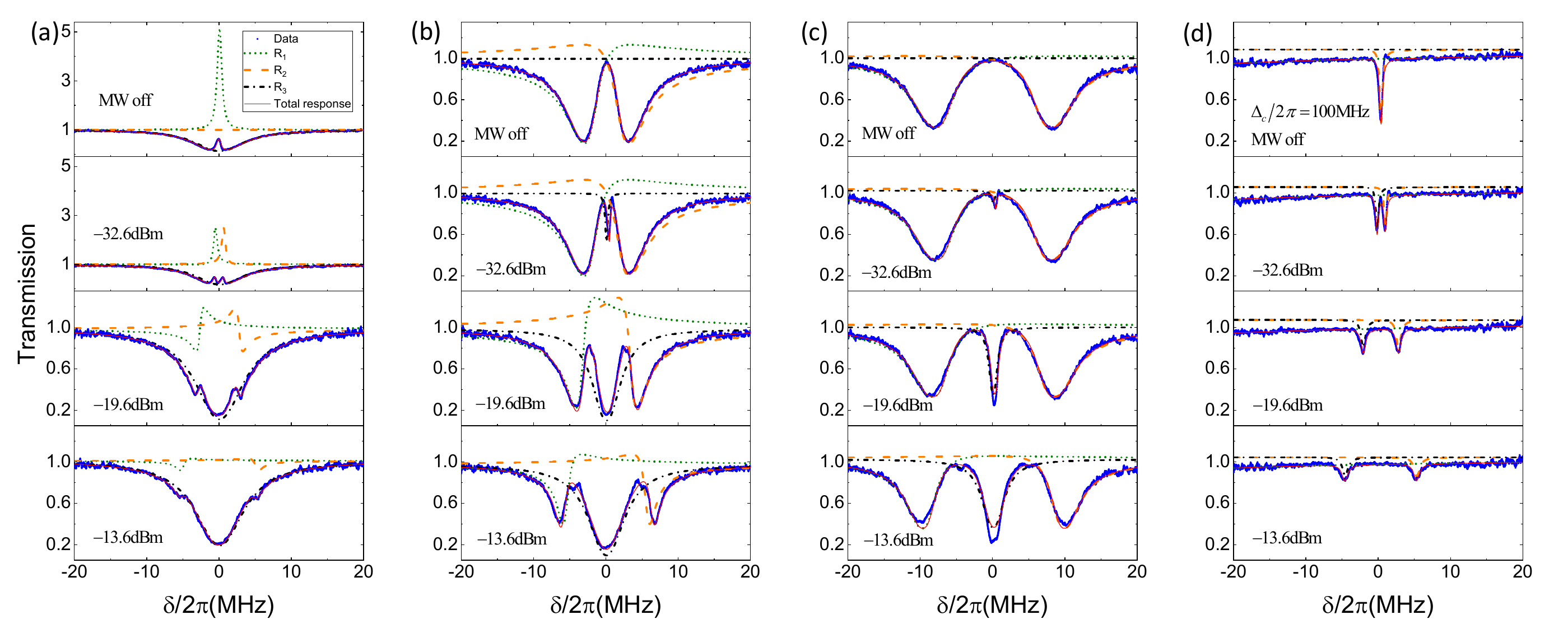}
\caption{ Normalized spectra of probe transmission for the indicated MW power input to the horn antenna. The four-level EIT spectra transition from DEIT to DATS is shown with the increase of coupling Rabi frequency: (a) DEIT with $\Omega_{c}/2\pi$ = 2 MHz, (b) crossover with $\Omega_{c}/2\pi$ = 6 MHz, and (c) DATS with $\Omega_{c}/2\pi$ = 16 MHz. (d)The EIA-ATS spectra with $\Delta_{c}/2\pi$ = 100 MHz, $\Omega_{c}/2\pi$ = 6 MHz, and $O\!D$ = 100. The blue dots show the experimental data, averaged over 1000 scans for each trace. Each result is fitted with the four-level susceptibility (red solid curves), and reveals three resonant responses $R_{1}$, $R_{2}$, and $R_{3}$ (green dotted, orange dashed, and black dash-dotted curves, respectively). From top to bottom, the MW Rabi frequencies are $2\pi\times$(0, 1, 5, 10) MHz, respectively. }
\end{center}
\end{figure*}

\section{Experimental results}

\subsection{Experimental results for four regimes}

 In our experiments, we measure the transmission $P_t$ via two-photon detuning $\delta$, and the results are shown with blue dots in Fig. 2. To analyze the data, we use Eq. (\ref{rho21}) with $O\!D$, $\Omega_{c}$, $\Delta_{\!M\!W}$, $\Omega_{\!M\!W}$, $\gamma_{3}$, and $\gamma_{4}$ as adjustable parameters to least-squares fit the data, while $\Delta_{c}$ and $\gamma_{2}\simeq\Gamma/2=2\pi\times 3$ MHz are fixed during the fits; the fitted results are plotted with red solid curves. We can retrieve these free parameters from the fits to reveal the three resonances $R_{1}$, $R_{2}$, and $R_{3}$; the results are shown in Fig. 2. These resonances demonstrate four regimes in the MW-driven four-level system: double EIT (DEIT) [$\Delta_c=0, \Omega_c<\Gamma$, Fig.2(a)], crossover [$\Delta_c=0, \Omega_c \approx\Gamma$, Fig.2(b)], double ATS (DATS) [$\Delta_c=0, \Omega_c>\Gamma$, Fig.2(c)] , and EIA ATS [$\Delta_c \gg \Omega_c,\Gamma$, Fig.2(d)].

Strong Fano interference occurs with a weak coupling laser ($\Omega_{c}\!<\!\Gamma$). The absorption profile in the top panel of Fig. 2(a) comprises two Lorentzians centered at the origin: One resonance is broad and positive and the other is narrow and negative. The MW field leads to a third transition pathway, and the destructive interferences among the three resonances induce two narrow transparent windows in other panels of Fig. 2(a), which indicates the presence of DEIT. When the coupling Rabi frequency is close to $\Gamma$, the interferences among three transition pathways manifest a crossover from DEIT to DATS, as shown in Fig. 2(b), and the constructive interferences start to prevail over destructive interferences. In the case of strong coupling, the rise of the third resonance $R_{3}$ between $R_{1}$ and $R_{2}$ in Fig. 2(c) demonstrates a transition from ATS to DATS. The resonances are approximate to three Lorentzians with the absence of destructive interference, and the separations between absorption peaks increase as the MW field strength increases. In the EIA ATS regime, as theoretically expected by Eq. (\ref{rho31}), the spectra in Fig. 2(d) are almost a double-Lorentzian function, and have a remarkable feature required for the SI-traceable measurement: The separation between the resonances $R_{2}$ and $R_{3}$ is exactly equal to $\Omega_{\!M\!W}$. The full width at half maximum of the EIA linewidth is about 400 kHz with $\Delta_c/2\pi$ = 100 MHz.

\begin{figure}[htb]
\begin{center}
\includegraphics[width=8.0cm]{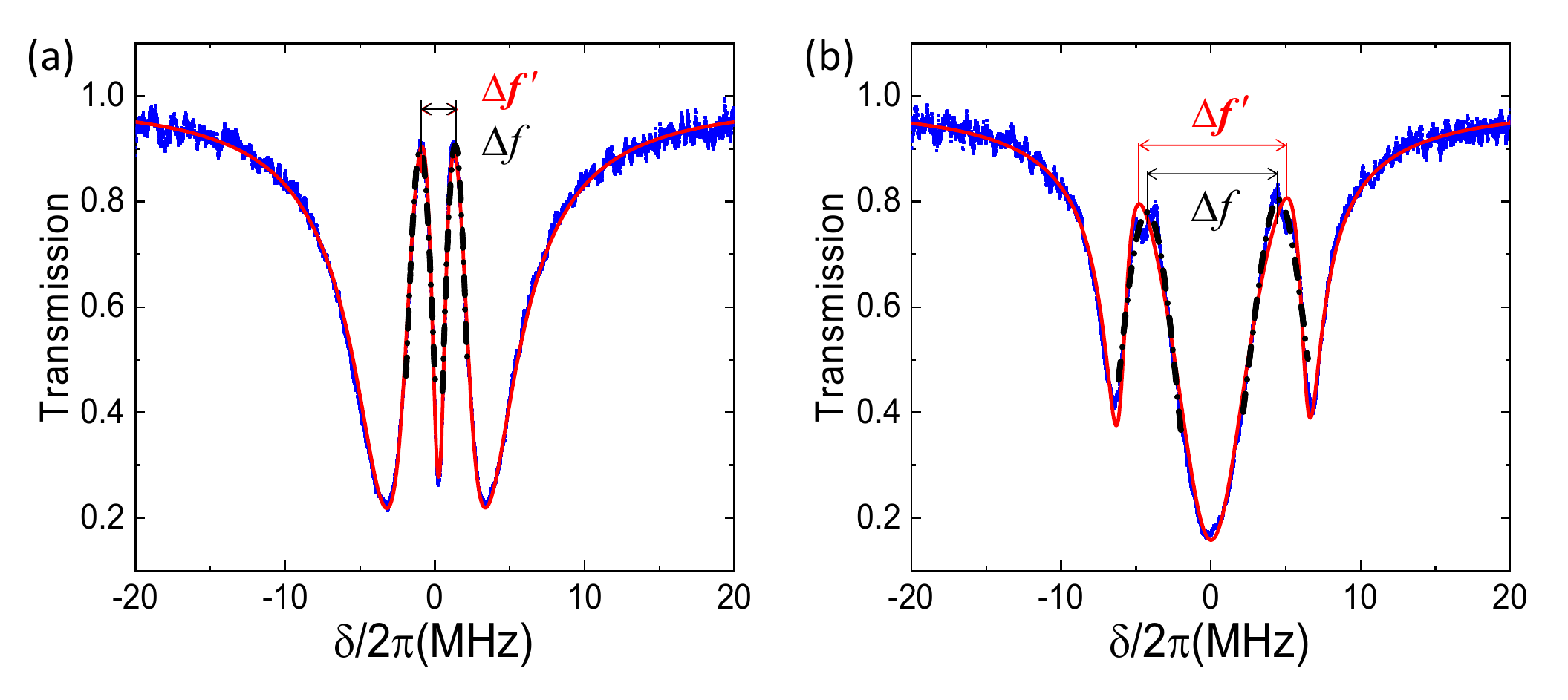}
\caption{ Spectral line fits for two cases in the crossover regime: ATS with MW power equal to (a) -26.6 dBm, and (b) -12.6 dBm. The ATS $\Delta\!f$ is extracted by locally fitting each peak with a Lorentzian function (black dash-dotted curve), while $\Delta\!f^{\prime}$ is extracted by globally fitting the whole spectrum with Eq.(\ref{rho21}) (red solid curve). }
\end{center}
\end{figure}

\subsection{THE AT SPLITTING AND ITS FITS}

\begin{figure*}[t]
\begin{center}
\includegraphics[width=13cm]{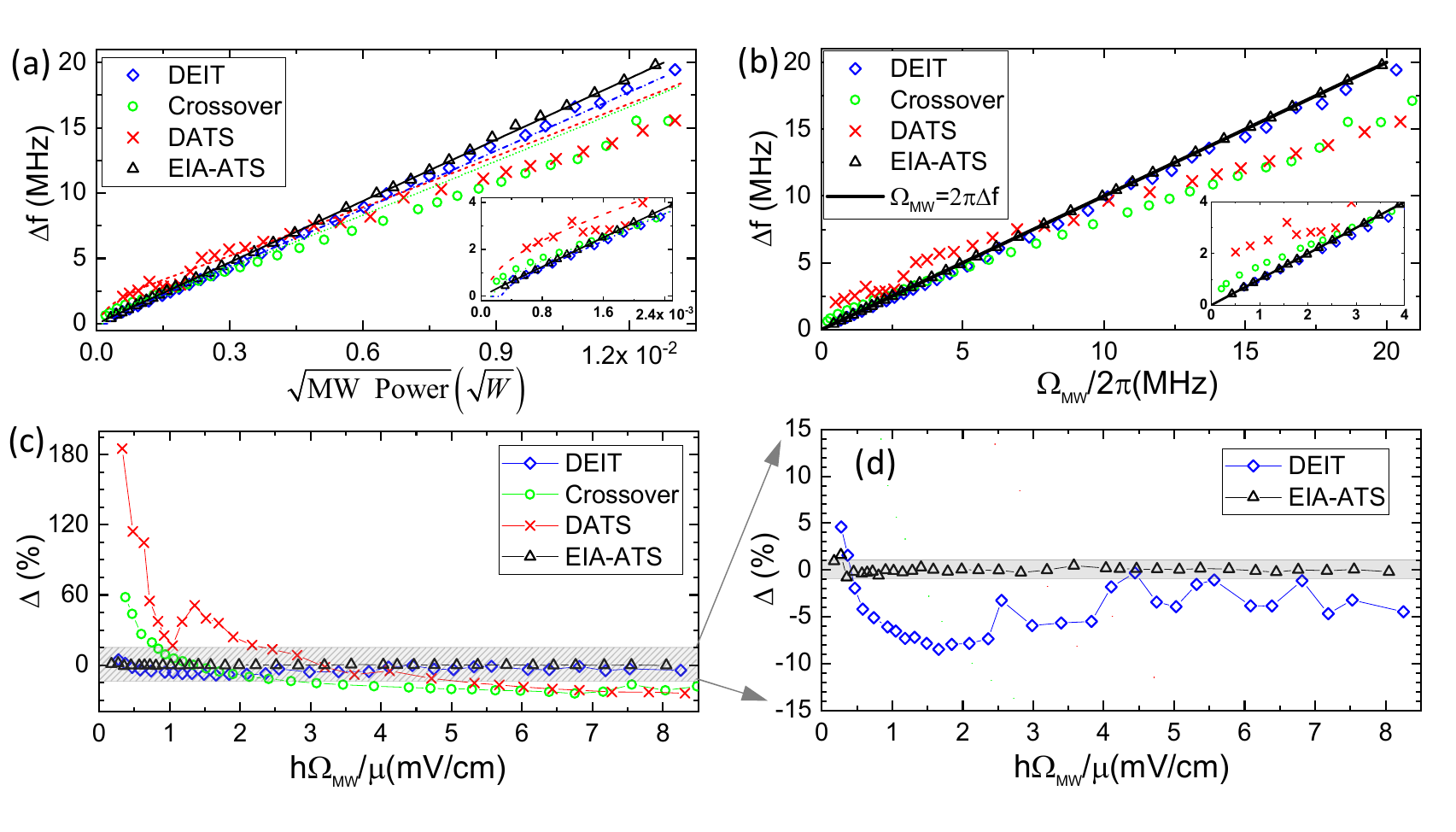}
\caption{ (a) The ATS $\Delta\!f$ as a function of MW power for the four regimes: DEIT, crossover, DATS and EIA ATS.
The symbols show the experimental data. The blue dash-dotted, green dotted, red dashed, and black solid lines are theoretically simulated from Eq. (2) for DEIT, crossover, DATS, and EIA-ATS, respectively.
(b) The ATS $\Delta\!f$ versus $\Omega_{\!M\!W}$ with the reference line $\Omega_{\!M\!W}= 2\pi\Delta\!f$.
The insets display a close-up of the corresponding parts. (c) Deviation $\Delta$ and (d) close-up of the shadow regime vs the MW-electric-field amplitude $|E|=\hbar\Omega_{MW}/\mu$.}
\end{center}
\end{figure*}

   The ATS is determined by the distance between two peaks in the spectrum.
To determine the distance, we need to find the extreme point in each peak, which is easy if the spectrum is a smooth function.  However, the measured spectrum is not a smooth function and there are fluctuations in the curves. To analyze the measured spectrum, we have two general methods to find the extreme point of the peak: global fit and local fit.
 We can use a Lorentzian function to locally fit a transmission peak ( based on least-squares method) and then we can  find the extreme point as well as the distance between two peaks. We can also use the polarization (susceptibility) given in Eq.(\ref{rho21}) to globally fit the whole measured spectrum ( based on the least-squares method).
The local fit with a Lorentzian function is usually used in the SI-traceable measurement (see Refs. \cite{shaffer1,Holloway2018}) because its traceability path is simple, direct, and independent of the numerous system variables, so it is just the goal of metrology organizations. On the other hand, the globally fitting method using the polarization given in Eq.(\ref{rho21}) is dependent on various adjustable parameters, such as $O\!D$, $\Omega_{c}$, $\Delta_{\!M\!W}$, $\Omega_{\!M\!W}$, and $\gamma_{3}$, $\gamma_{4}$, that is, the traceability path using Eq. (\ref{rho21}) to retrieve the splitting is indirect, complex, and relies on the calibration of numerous experimental parameters.

Typical examples of the fits are shown in Fig. 3, where we denote by $\Delta f$ the ATS fitted with a local Lorentzian function and by $\Delta\!f^{\prime}$ the splitting fitted with a global polarization given in Eq. (\ref{rho21}). They can be the same or different, depending on the detailed control parameters. Figure 3 shows that the whole spectrum is very similar to the theoretical calculation using Eq.(\ref{rho21}), but locally they are different  because of the local fluctuations in the measured spectrum. As shown in Fig. 3(a), both the Lorentzian function and polarization function fit very well to the transmission peaks. With the increase of MW power, the matching degree of the polarization function with the peaks around their locations decreases but the Lorentzian function still fits well at the peak location as shown in Fig. 3 (b). Under this condition, it is more reliable to retrieve a global variable such as $\Omega_{\!M\!W}$ than the extraction of a local quantity of the ATS $\Delta\!f^{\prime}$ from the polarization fit. In addition, there are two other reasons that the Lorentzian fit is effective and essential. First, its traceability path is direct, as we have mentioned. Second, $\Omega_{\!M\!W}$ and $\Delta\!f$ are respectively derived from the completely independent fitting methods, and thus the electric-field deviation can be reliably and reasonably evaluated in the SI-traceable measurement.

\subsection{Relations between the AT splitting and MW Rabi frequency}

We now evaluate the accuracy of the measurement if Eq.(1) is used. The ATS $\Delta f$ fitted with the Lorentzian function  as a function of MW power input to the horn antenna are shown in Fig.4(a).
In Fig. 4(a) the large departures of the extracted $\Delta\!f$ from the theoretical curves in the crossover and DATS regimes are due to the broad and asymmetric peaks not being of Lorentzian form. In contrast, the experimental data in the DEIT and EIA ATS regimes agree well with the theoretical simulations.
As shown in Fig. 4(a), the nonlinear behavior always happens at the beginning of ATS in the EIT scheme.
In contrast, the nonlinear behavior almost disappears in the EIA ATS regime as a result of the adiabatic elimination of the excited state.

To examine the equivalence relation given in Eq. (\ref{Rabi_AT}), 
we retrieve the MW Rabi frequencies $\Omega_{\!M\!W}$ from the fitted spectra (red solid curves in Fig. 2) using Eq. (\ref{rho21}) and then plot the relations between $\Delta\!f$ and $\Omega_{\!M\!W}/2\pi$ in Fig. 4(b) for the data shown in Fig. 4(a).  It shows that the relation in Eq. (\ref{Rabi_AT})  breaks down in the EIT condition. In sharp contrast, the data in the EIA ATS regime and the reference line perfectly coincide with each other, as expected from Eq. (\ref{rho31}).
To further quantitatively characterize the measuring accuracy,  we calculate the percent deviation $\Delta$ between the measured splitting and $\Omega_{\!M\!W}$ for the data of Fig. 4(b), defined by
\begin{equation}
\label{Delta}
\Delta=100\times(2\pi\Delta\!f-\Omega_{MW})/\Omega_{MW}.
\end{equation}
The results are plotted in Figs. 4(c) and (d), which show that the deviations for DEIT, crossover, and DATS are generally larger than $5\%$. So the measured amplitudes would be either greatly underestimated or greatly overestimated if the relation (1) is used in experiments. In contrast, the deviation for our EIA method is less than $1\%$ for $|E|\geqslant$350 $\mu \mathrm{V} \mathrm{cm}^{-1}$. It slightly increases to $1.0\%$ ($1.6\%$) for $|E|\approx 178.8$ ($271.7) \mu \mathrm{V} \mathrm{cm}^{-1}$. In the later, we will show that the detected field can be reduced to around 100 $\mu \mathrm{V} \mathrm{cm}^{-1}$ when the coupling $\Delta_c/2\pi$ increases to 200 MHz.

\begin{figure*}[htb]
\begin{center}
\includegraphics[width=13cm]{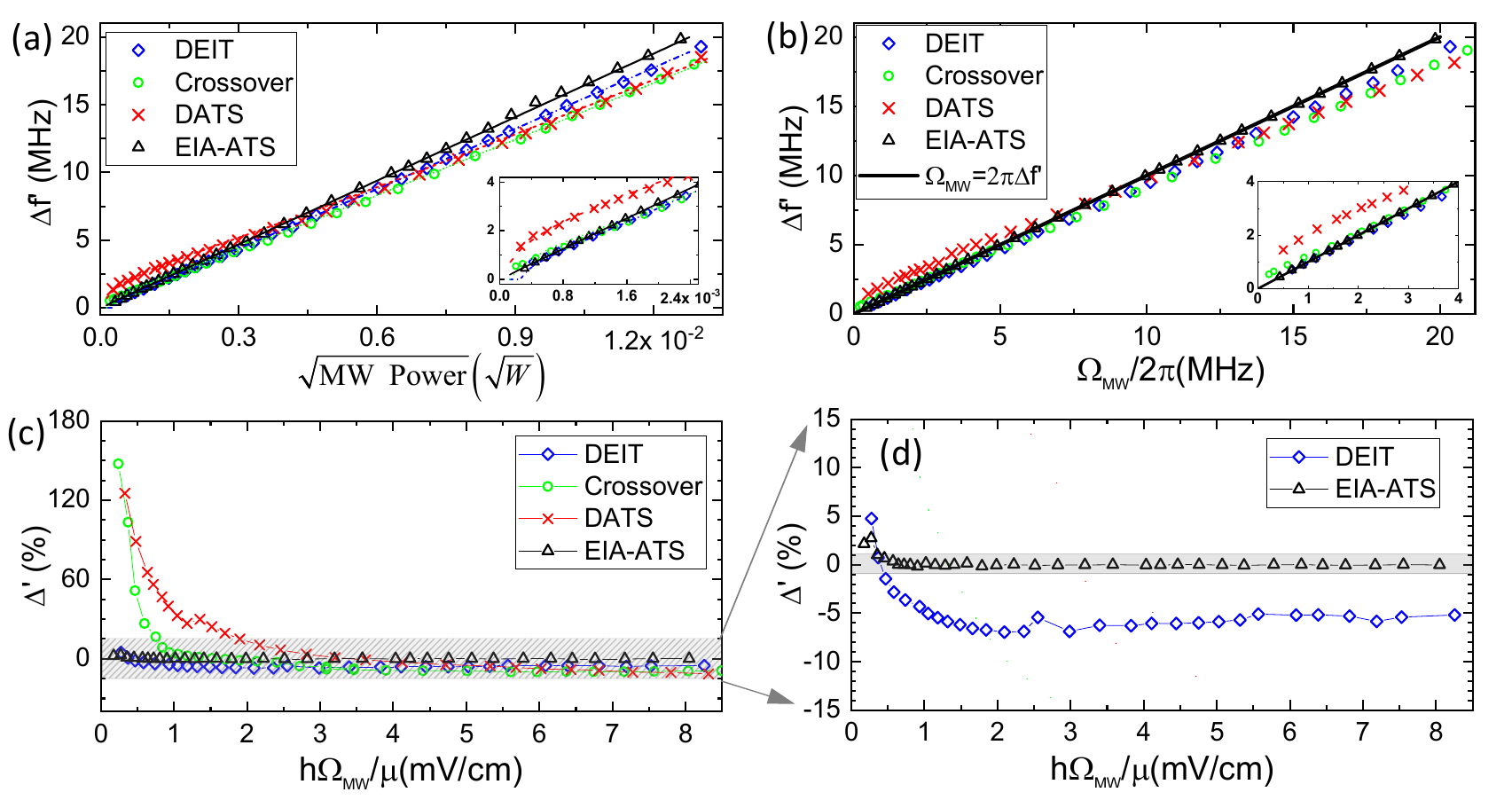}
\caption{(a) The ATS $\Delta\!f^{\prime}$ as a function of MW power for the four regimes.
The symbols show the experimental data. The blue dash-dotted, green dotted, red dashed, and black solid lines are the theoretically simulated curves for DEIT, crossover, DATS, and EIA ATS regions, respectively, which are obtained by finding the extreme point of each peak in the four-level susceptibility. (b) The ATS $\Delta\!f^{\prime}$ versus $\Omega_{\!M\!W}$ with the $\Omega_{\!M\!W}= 2\pi\Delta\!f^{\prime}$ reference line.
The insets display a close-up of the corresponding parts. (c) Deviation $\Delta^{\prime}$ and (d) close-up of the shadow region vs the MW-electric-field amplitude $|E|=h\Omega_{\!M\!W}/\mu.$ }
\end{center}
\end{figure*}

\begin{figure}[htb]
\begin{center}
\includegraphics[width=8cm]{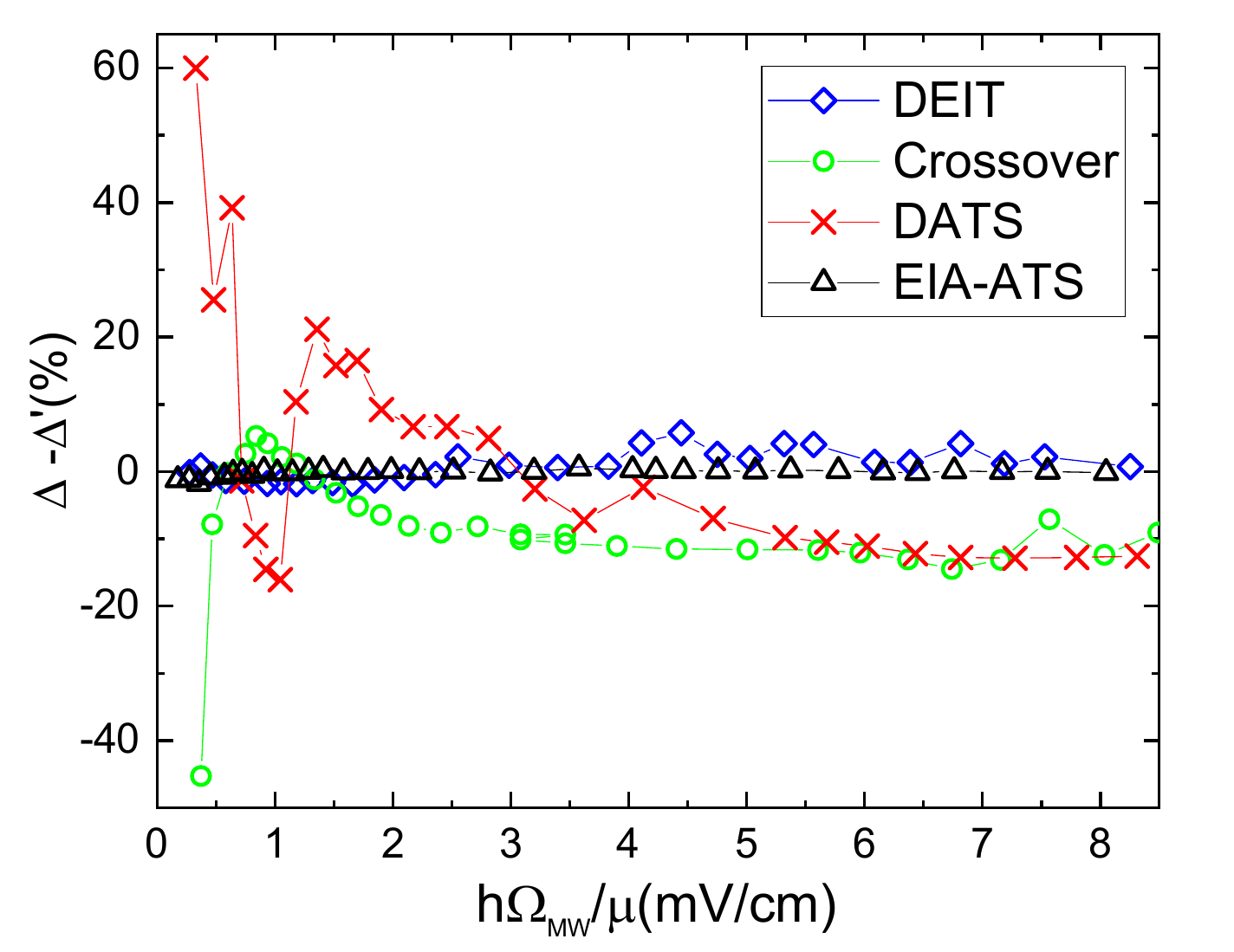}
\caption{Differences $\Delta-\Delta^{\prime}$ vs the MW-electric-field amplitude. }
\end{center}
\end{figure}

We also analyze the features of the ATSs $\Delta\!f^{\prime}$ derived from the polarization fits to the data of Fig. 4(a). Figure 5(a) shows that all measured data in four regimes agree well with the theoretical curves. In the polarization fits, we can also derive the $\Omega_{\!M\!W}$ and thus we plot  the relations between $\Delta\!f^{\prime}$ and $\Omega_{\!M\!W}/2\pi$ in Fig. 5(b). To quantitatively characterize the errors used in Eq.(1) to extract the electric field, we introduce the percent deviation $\Delta^{\prime}$ between the splitting $\Delta\!f^{\prime}$ and $\Omega_{\!M\!W}$ for the data of Fig. 5(b), defined by
\begin{equation}
\label{DeltaP}
\Delta^{\prime}=100\times(2\pi\Delta\!f^{\prime}-\Omega_{MW})/\Omega_{MW}.
\end{equation}
The results are plotted in Figs. 5(c) and (d), which show that the deviations for DEIT, crossover and DATS are generally larger than $7\%$. Thus the measured MW-electric-field amplitudes would be either greatly underestimated or greatly overestimated if the relation (1) is used in experiments. In contrast, the deviation for our EIA method is less than $1\%$ for $|E|\geqslant$350 $\mu \mathrm{V} \mathrm{cm}^{-1}$. It slightly increases to around $3.0\%$  for $|E|\approx 178.8$ and $271.7 \mu \mathrm{V} \mathrm{cm}^{-1}$.

Comparing the results in Figs. 4 and 5, we find that the detailed results may be different for the two fitting methods, but the main conclusions for the four regimes explored in this paper are independent of the fitting methods. Thus the deviations defined in Eq.(\ref{Delta})  can be used to characterize the intrinsic errors.

We can further quantitatively compare the two fitting methods by calculation of the difference $\Delta-\Delta^{\prime}$; the results are plotted in Fig. 6, which shows that the differences $\Delta-\Delta^{\prime}$ of the two fitting methods are less than $1\%$ for EIA regime, but are generally relatively large for EIT regimes, especially in the crossover and DATS regimes. The main reasons are due to the broad and asymmetric peaks in the latter regimes.

\subsection{Performances of the EIA method}

\begin{figure*}[htb]
\begin{center}
\includegraphics[width=13cm]{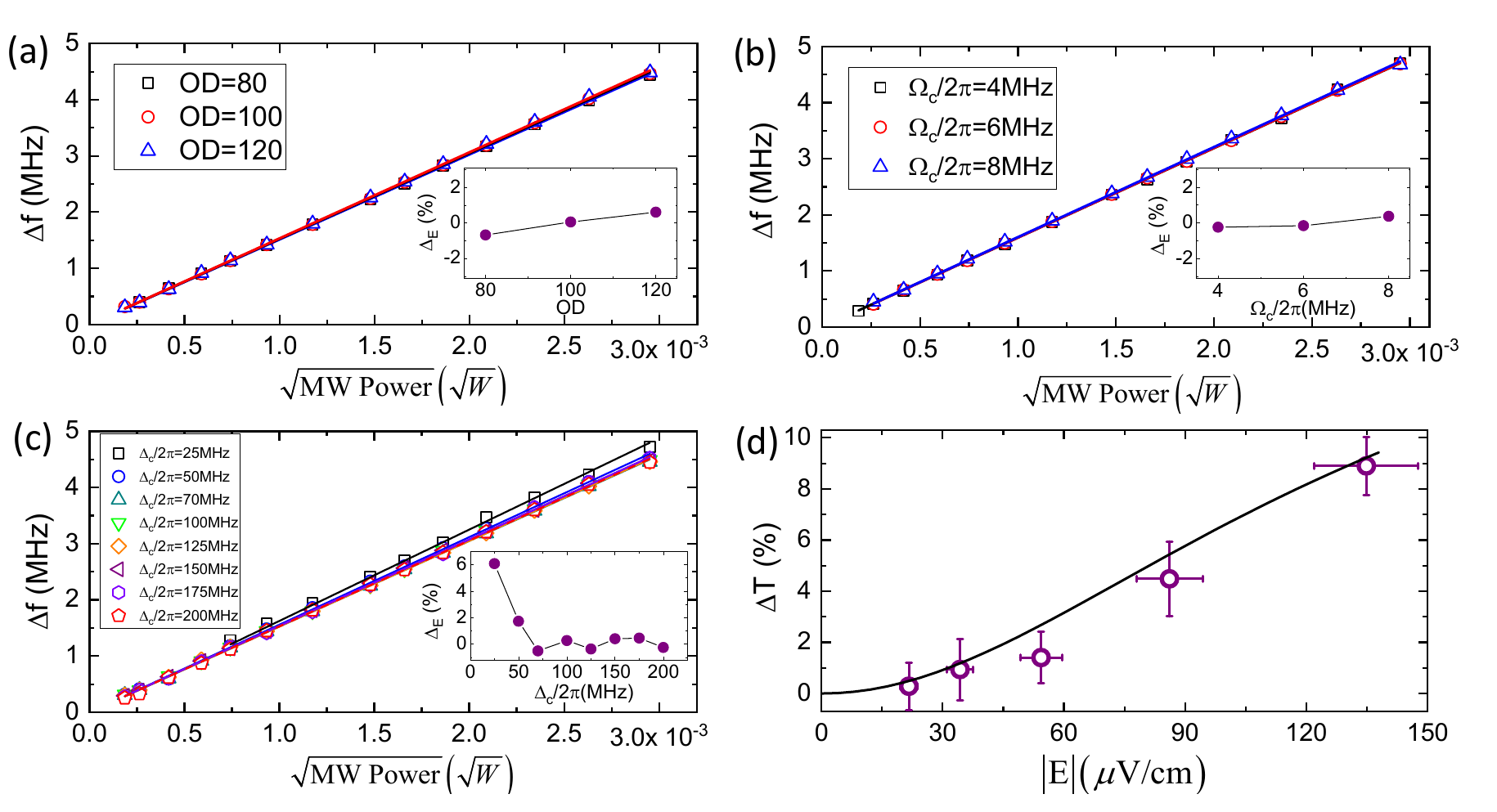}
\caption{ The EIA ATS $\Delta\!f$ as a function of MW power for the indicated values of (a) optical depths, (b) coupling Rabi frequencies, and (c) coupling detunings. (d) Transmission difference $\Delta T$ in EIA signal vs MW-electric-field amplitude. The symbols represent experimental data. The curves in (a)-(c) are linear fitting results, while the solid line in (d) is obtained from Eq.(\ref{rho21}) with $O\!D$ = 70, $\Delta_c/2\pi$ = 100 MHz, and $\Omega_c/2\pi$ = 6 MHz. The vertical error bars are the standard deviation of five measurements and the horizontal error bars correspond to the estimated uncertainties of the output power of the MW generator. The insets show the relative deviations of the MW electric field $\Delta_{E}$.}
\end{center}
\end{figure*}

  We further analyze the dependence of the EIA ATS on optical depth, Rabi frequency, and detuning of the coupling field. Except for the indicated variables, the EIA measurement is taken under the same conditions as in Fig. 2(d). Figure 7 shows that the linear relations between EIA ATS and the applied MW electric field hold well for different control parameters (i.e., $O\!D$, $\Delta_c$ and $\Omega_c$). The figure also indicates the relative deviation $\Delta_{E}$ between the measured MW-electric-field strength (which is proportional to the splittings) and the applied MW electric field. As the fitted lines perfectly pass through the original point, $\Delta_{E}$ is equal to the difference between the slope coefficients retrieved from linear fits and calculated with Eq. (1).  Figures 7(a) and (b) show that all $\Delta_{E}$ are below 1\%, and thus the EIA measurements are very robust against the variations of the optical depth and $\Omega_c$.
Figure 7(c) shows that, as long as the excited state $|2\rangle$ is strongly detuned ($\Delta_c>$10 $\Gamma$), the fitted lines start to converge and the deviations $\Delta_{E}$ are below 1\%. Here an ATS as small as $\sim$250 kHz is observed at $\Delta_c/2\pi$ = 200 MHz, and the corresponding MW electric field is 101.4 $\mu\mathrm{V}\mathrm{cm}^{-1}$, which is the smallest SI-traceable field strength we have detected.  Note that several $m\mathrm{V} \mathrm{cm}^{-1}$ for the MW electric field can be measured with room-temperature vapor cells and $1.2 mV cm^{-1}$ is achieved for the static electric field for a 300 $\mu s$ detection time ( corresponding to $30 \mu V cm^{-1} Hz^{-1/2}$ )\cite{cat}.
 At present, the smallest detected MW electric field is limited by the Rydberg level broadening induced by background electric and magnetic fields and the technical noise from detection and laser intensity fluctuations. While there is a lower limit set by the laser linewidth, spontaneous decay, interaction time, etc., a lower limit for the traceable MW-electric-field measurement in the range $\lesssim$10 $\mu\mathrm{V}\mathrm{cm}^{-1}$ seems quite feasible, considering that the present dephasing rates $\gamma_3$ and $\gamma_4$ are ten times greater than the corresponding Rydberg decay rates (about 18 kHz under room temperature blackbody radiation).

To achieve a continuous MW-electric-field measurement from the EIA linear region to the nonlinear region, we scan the probe laser frequency at low MW field and determine its transmission difference $\Delta T$ relative to the three-level EIA signal. Figure 7(d) shows the percent difference $\Delta T$ as a function of the MW-electric-field amplitude. In the sub-ATS region, the EIA signal smoothly decays as the electric-field amplitude increases. The dip of the sub-ATS is most similar to a single Lorentzian in shape, and its depth is extracted from fitting a Lorentzian function to the spectra.
The solid curve shows the theoretical calculation with four-level susceptibility, which agrees with the experimental data. The smallest detectable strength in Fig. 7(d) is 21.6 $\pm$ 2.1 $\mu \mathrm{V} \mathrm{cm}^{-1}$, mainly limited by the intensity stability of the probe laser. Note that the intrinsic transmission change in our MW electrometry is about ten times greater than that of the EIT signal in vapor cells \cite{shaffer1}, which implies that the lower limit of the smallest detectable MW field can be achieved.


\section{Discussion and conclusion}

\begin{figure*}[htb]
\begin{center}
\includegraphics[width=15.5cm]{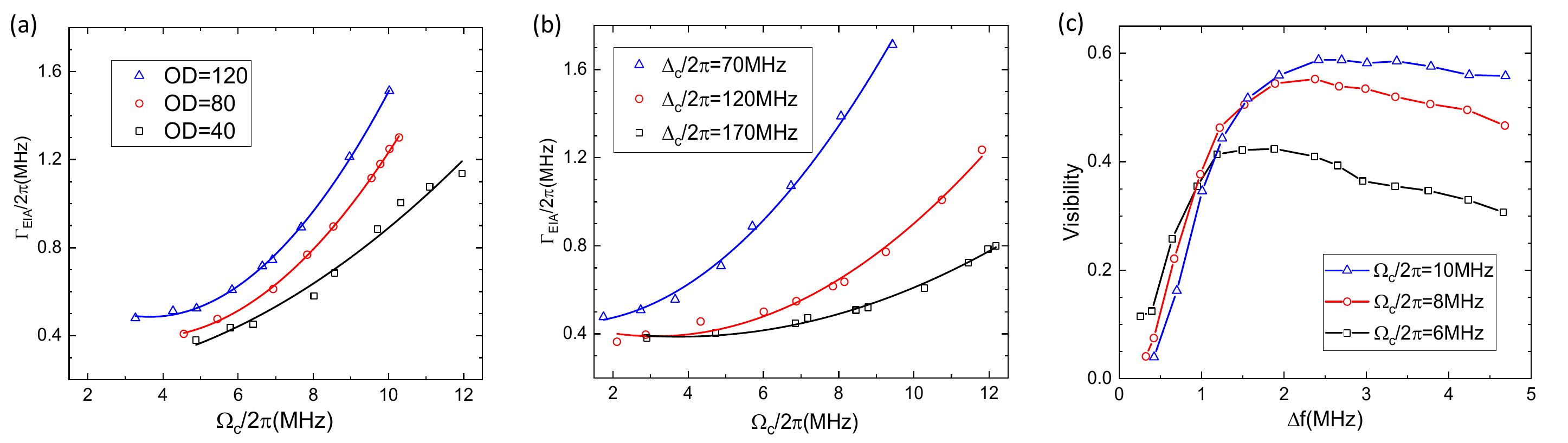}
\caption{ The EIA linewidth $\Gamma_{\!E\!I\!A}$ versus $\Omega_{c}$ for the indicated values of (a) optical depths and (b) coupling detunings. The curves are the quadratic fitting results. (c) Visibility $V$ of the EIA ATS vs $\Delta f$ for the indicated coupling Rabi frequencies. The symbols in each figure show the experimental data. The measurement is taken under the same conditions as in Fig. 2(d), except for the indicated variables.}
\end{center}
\end{figure*}

We now briefly discuss the EIA linewidth and visibility of EIA ATS. The range and resolution of the SI-traceable measurement of the MW-electric-field strength are related to the linewidth of the EIA signal. Figure 8 shows the dependence of the EIA linewidth on the optical depth, the Rabi frequency, and detuning of coupling field. Its full width at half maximum is extracted by fitting Eq. (\ref{rho21}) to the spectra for $\Omega_{\!M\!W}$ = 0. The EIA profile is approximate to the absorption profile of an effective two-level system. In cold atom, there is no Doppler factor and the EIA linewidth follows $\Gamma_{\!E\!I\!A} \simeq (\sqrt{O\!D} \Omega_{c}^{2})/ (8\Delta_{c})$. Thus $\Gamma_{\!E\!I\!A}$ decreases as $O\!D$ and $\Omega_{c}$ decrease (or as $\Delta_{c}$ increases). In principle, the small linewidth allows the direct SI-traceable measurement of a very weak MW field, which is just limited by the laser and Rydberg-level linewidth.

Figure 8(c) shows the dependence of the visibility of EIA ATS on $\Delta f$ at a different coupling Rabi frequency $\Omega_{c}$. The visibility is defined as $V\!=\!(\alpha_{max}-\alpha_{min})/\alpha_{bg}$, where $\alpha_{min}$ and $\alpha_{max}$ are the minimum transmission coefficient at the bottom of the EIA dip and the maximum coefficients between the two EIA dips, respectively, and $\alpha_{bg}$ is the background transmission coefficient. As $\Omega_{c}$ decreases, the minimum detectable EIA ATS reduces due to the narrower linewidth, and meanwhile the visibility of the AT-splitting decreases due to the shallower EIA dip, which is a trade-off for the weak MW-electric-field measurement. The visibility $V$ decreases with the increase of $\Omega_{\!M\!W}$ when the ATS is larger than $2\Gamma_{\!E\!I\!A}$ . This phenomenon may arise from the Rydberg dephasing enhanced by the resonant dipole-dipole interaction \cite{dp}, but the detailed effects of the interactions  are beyond the scope of this work.

In summary, we have demonstrated a sensitive method for direct SI-traceable measurements of the MW electric field based on EIA in cold Rydberg atoms, which shows clear advantages serving as a traceable standard for MW electrometry.
At present, the detection is not shot-noise limited and the apparatus can be improved in many ways \cite{sr, oe}, including the use of lasers with narrower linewidth and lower amplitude noise, the use of lower noise detectors, the implementation of a homodyne detection, frequency modulation spectroscopy  or a Schr\"odinger-cat state based measurement \cite{cat}. In combination with these technologies, a much lower SI-traceable MW-electric-field measurement can be achieved with this EIA method.


\vspace{0.5cm}
\begin{acknowledgments}
 The authors thank Zhenfei Song for useful discussions. This work was supported by  the National Natural Science Foundation of China (Grants No. 91636218, No. 11822403, No. 11804104,  No. 61875060 and No. U1801661), the Key-Area Research and Development Program of GuangDong Province (Grant No. 2019B030330001), the National Key Research and Development Program of China (Grants No. 2016YFA0301803 and No. 2016YFA0302800),  the Natural Science Foundation of Guangdong Province (Grants No. 2018A030313342 and No. 2018A0303130066), and the Key Project of Science and Technology of Guangzhou (Grants No. 201804020055 and No. 201902020002).

K.-Y.L. and H.-T.T. contributed equally to this work.
\end{acknowledgments}

\begin{appendix}

\section{Uncertainties}

\begin{table}[h]
\caption {\small {Main sources of uncertainty in the present measurement.}}
\centering
\label{table:T1}
\begin{ruledtabular}
\begin{tabular}{ccc}
\makecell*[l]{Effect} &  \makecell*[l]{EIA-AT\\ splitting} & \makecell*[l]{EIA\\ transmission}  \\
\colrule
\makecell*[l]{Stray magnetic \\field$^{\rm a}$ ($\sim$10 mG)} & \makecell*[l]{0.1\%} &\makecell*[l]{0.1\%}  \\
\makecell*[l]{Stray electric field$^{\rm a}$ \\($\sim$1{m\!V}{cm}$^{-1}$)} & \makecell*[l]{0.005\%} & \makecell*[l]{0.005\%}  \\
\makecell*[l]{Optical depth change\\($\Delta{O\!D}^{\rm b}\sim$5\%)} & \makecell*[l]{${\rm c}$} &\makecell*[l]{2\%}  \\
\makecell*[l]{Microwave source \\frequency} & \makecell*[l]{0.42\%} & \makecell*[l]{1.2\%} \\
\makecell*[l]{Microwave source \\amplitude$^{\rm d}$} & \makecell*[l]{${\rm e}$} & \makecell*[l]{4.2\%} \\
\makecell*[l]{Two-photon detuning\\noise ($\sim$30 kHz)} & \makecell*[l]{${\rm f}$} & \makecell*[l]{0.1\%} \\
\makecell*[l]{Technical noise from \\ laser intensity noise \\and detection} & \makecell*[l]{1\%} & \makecell*[l]{$\sim$3-9\% for $\Delta T$,\\ growing from largest\\to smallest signal} \\
\end{tabular}
\end{ruledtabular}
\begin{flushleft}
\footnotesize{$^{\rm a}$See Ref. \cite{shaffer1}.}\\
\footnotesize{$^{\rm b}$$\Delta {O\!D}$ is the change in optical depth during a measurement.}\\
\footnotesize{$^{\rm c}$This changes absorption but does not change splitting.}\\
\footnotesize{$^{\rm d}$Related to the amplitude uncertainty at low MW field.}\\
\footnotesize{$^{\rm e}$This is calibrated with EIA ATS.}\\
\footnotesize{$^{\rm f}$This increases $\Gamma_{\!E\!I\!A}$ but does not change splitting at this noise level.}\\
\end{flushleft}
\end{table}

The principal sources of systematic uncertainties and statistical noises in the EIA ATS and peak transmission measurements are listed in Table 1.
The dominant sources of systematic uncertainties in our present experiment are the change of dipole moment arising from background magnetic field, the change in optical depth, and the uncertainties associated with the microwave generator. The system uncertainty due to the stray magnetic field can be reduced by shielding the cell or by dynamically compensating for the magnetic field. The principal source of statistical uncertainties in both experiments is the technical noise associated with the frequency and intensity instability of the probe and coupling lasers. The smallest detected EIA ATS and change in transmission are mainly limited by the technical noises from laser intensity noise and detection. At present the detection is not shot-noise limited and the apparatus can be improved, by reducing the laser intensity noise. To apply Rydberg-atom-based MW detection in practice, the MW field variation arising from the Fabry-Perot effect of the vacuum-cell should be minimized by making the vacuum cell size small compared to the MW wavelength \cite{cell}.

\end{appendix}

\newpage


\end{document}